\documentclass[twocolumn,prl,showpacs,floatfix]{revtex4}
\usepackage{amssymb}

\usepackage{amsmath}
\usepackage{epsfig}
\usepackage{graphics}

\begin{document}

\title{Rashba spin-orbit interaction and shot noise for spin-polarized and
entangled electrons}
\author{J. Carlos \surname{Egues}}
\altaffiliation[Permanent address: ]{Department of Physics and Informatics, University of S\~ao Paulo at S\~ao
Carlos, 13560-970 S\~ao Carlos/SP, Brazil.}
\author{Guido Burkard}
\author{Daniel Loss}
\affiliation{Department of Physics and Astronomy, University of
Basel, Klingelbergstrasse 82, CH-4056 Basel, Switzerland}

\begin{abstract}
We study shot noise for spin-polarized currents and entangled
electron pairs in a four-probe (beam splitter) geometry with a
local Rashba spin-orbit (s-o) interaction in the incoming leads.
Within the scattering formalism we find that shot noise exhibits
Rashba-induced oscillations with continuous bunching and
antibunching. We show that entangled states and triplet states can
be identified via their Rashba phase in noise measurements. For
two-channel leads we find an additional spin rotation due to s-o
induced interband coupling which enhances spin control. We show
that the s-o interaction determines the Fano factor which provides
a direct way to measure the Rashba coupling constant via noise.
\end{abstract}

\date{\today}
\maketitle


Spin-related effects in transport form the basis of the emerging field of
semiconductor spintronics \cite{springer}. Moreover, the \emph{electric}
control of intrinsic \emph{magnetic} degrees of freedom offers an important
mechanism to manipulate and probe spin transport. For instance, the
spin-transistor proposal of Datta and Das \cite{datta} highlights the
relevance of a gate-controlled Rashba spin-orbit interaction as a means of
spin rotating electron states in one-dimensional channels.

\begin{figure}[h]
\begin{center}
\epsfig{file=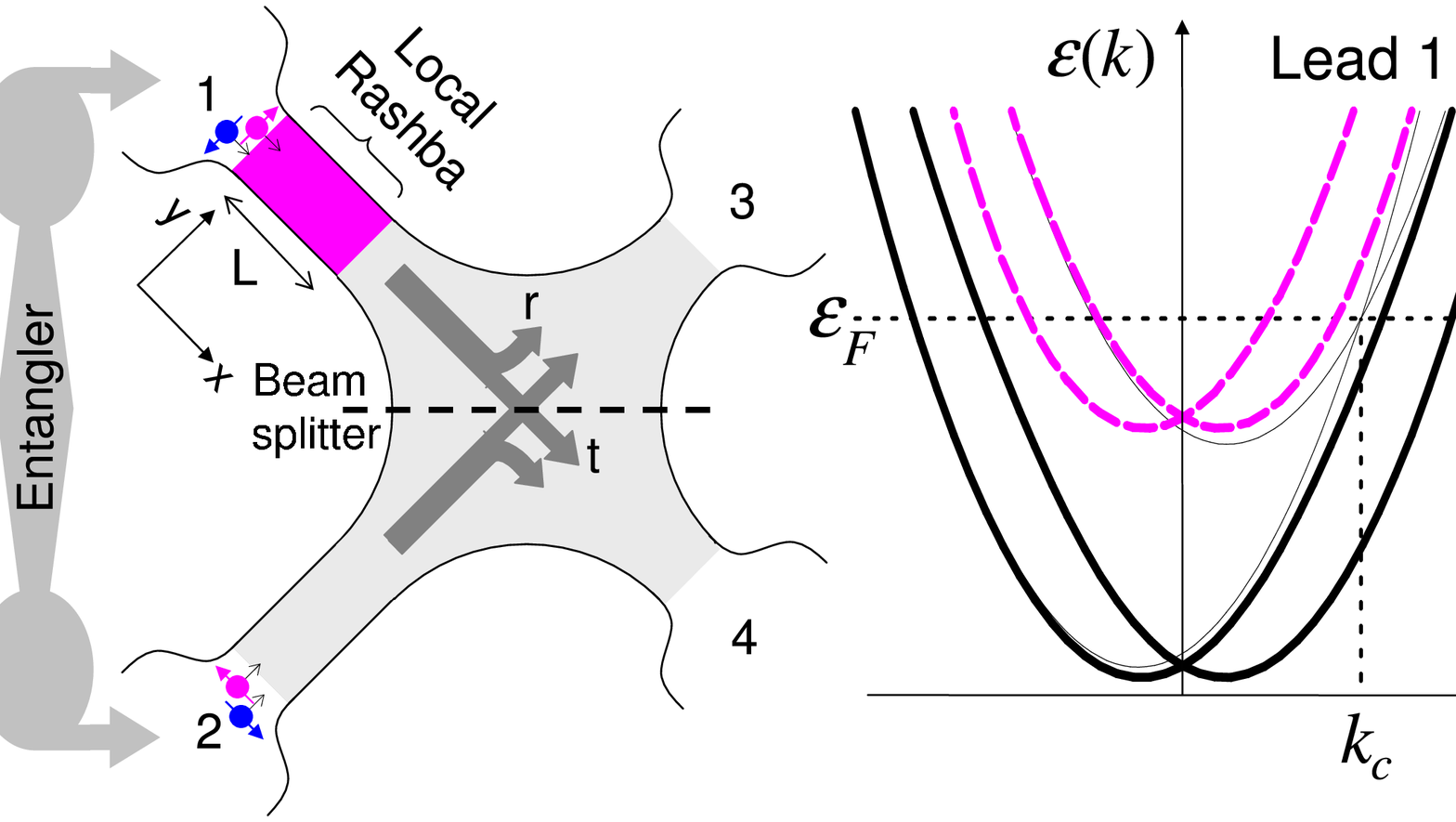,width=0.45\textwidth}
\end{center}
\par
\vspace{-5mm}
\caption{Left: beam-splitter geometry with a local Rashba s-o interaction in
lead 1. Electron pairs (entangled or not) are injected into leads 1 and 2.
Electrons in lead 1 undergo a phase shift (spin rotation). This Rashba phase
\emph{continuously} changes the symmetry of the \emph{spin} part of the pair
wave function and induces sizable oscillations in the noise for both
spin-polarized and entangled electrons. Right: Rashba bands with (thick
solid and dashed lines) and without (thin solid lines) s-o interband
coupling. Interband coupling yields band anti crossing and induces an
additional spin rotation for electrons impinging near the band crossing
point $k_{c}$.}
\label{fig:fig1}
\end{figure}

In this work, we investigate the transport properties of spin-polarized \cite%
{spin-pol,egues} and spin-entangled \cite{entangler,BLS} electrons for a
beam-splitter configuration \cite{liu,oliver,henny} with a local Rashba
spin-orbit (s-o) interaction, which acts within a finite region (length $L$)
of the incoming leads, see Fig.~\ref{fig:fig1} (left). Due to such a local
s-o term the spinors of the incoming electrons can be varied continuously,
which affects then the orbital symmetry of the wave function (via the Pauli
principle) and thus transport properties such as current and noise. Within
the scattering approach \cite{buttiker} we calculate current noise for leads
with one and with two channels, see Fig. 1 (right). We find that shot noise
for spin-polarized and entangled electrons strongly oscillates as a function
of the Rashba coupling and the length $L$. In particular, singlet (triplet)
pairs exhibit \emph{intermediate} degrees of (anti-)bunching behavior \cite%
{BLS,taddei}. We show that entangled pairs as well as triplet states can be
identified in noise experiments via their Rashba phase. We find that the s-o
interaction determines the Fano factor, implying that the Rashba coupling
can be measured via noise. Finally, we find an additional spin phase due to
s-o induced interband coupling for leads with two channels. Since this
modulation can be varied via the lateral confinement of the lead, this
effect provides a new mechanism for electrical spin control.

\emph{System.} We consider an experimentally feasible beam-splitter geometry %
\cite{liu,oliver,henny} with two incoming and two outgoing leads, Fig.~\ref%
{fig:fig1}. We assume that the local Rashba coupling in this lead can be
externally controlled via a proper gating structure \cite{nitta,grundler}.
The injected electrons (entangled or not) undergo a local spin evolution
within an extension $L$ in lead 1. Note that we do not consider a Rashba
interaction in lead 2; we are interested only in \emph{phase differences}
between leads 1 and 2. However, it is straightforward to extend our analysis
to include s-o interaction in both incoming leads. We consider first
single-channel leads (one occupied band), and then move on to the more
involved case of two channels. For an incoming plane wave with wave vector
\emph{k} along the \emph{x} direction the Rashba term is simply $%
H_{R}=-\alpha k\sigma _{y}$ \cite{rashba,molenkamp}, where $\alpha $ denotes
the s-o coupling constant.

\emph{Spin transfer operator.} The rotation of the spin state can be
described by a unitary transfer operator. This unitary operator can then be
incorporated straightforwardly into the usual scattering matrix (Landauer-B%
\"{u}ttiker) formalism for coherent transport \cite{buttiker} in order to
calculate the current and the current correlators. We find that the transfer
operator through the Rashba region is $U_{R}=\exp \left( -i\theta _{R}\sigma
_{y}/2\right) $, i.e., a rotation about the $y$ axis by an angle $\theta
_{R}=2\alpha m^{\ast }L/\hbar ^{2}=2k_{R}L$, $m^{\ast }$ denotes the
electron effective mass. This Rashba rotation is well known \cite{datta} and
occurs only for incoming electrons having a spin component perpendicular to $%
y$. In the basis of the eigenstates of $\sigma _{z}$, we can write
\begin{equation}
U_{R}=\left(
\begin{array}{cc}
\cos \theta _{R}/2 & -\sin \theta _{R}/2 \\
\sin \theta _{R}/2 & \cos \theta _{R}/2%
\end{array}%
\right) .  \label{eq2}
\end{equation}%
For instance, an incoming plane wave in lead 1 with spin up, $|\psi _{%
\mathrm{in}}\rangle =|k\uparrow \rangle $, emerges at the other side of the
Rashba region in the rotated state \cite{dynamical}
\begin{equation}
\langle x=L|\psi _{\mathrm{out}}\rangle =\left(
\begin{array}{c}
\cos \theta _{R}/2 \\
\sin \theta _{R}/2%
\end{array}%
\right) .  \label{eq5}
\end{equation}
In the above we assume a \emph{unity} transmission probability $T_{R}$ for
electrons through the Rashba region. Indeed, $T_{R}\approx 1$ since there is
no additional band offset due to different materials in our incoming lead %
\cite{mismatch}. This implies that the Rashba interaction does not directly
introduce noise in the lead -- it simply rotates the incoming spin state.
However, indirectly it does affect the noise characteristic of the entire
system since it effectively changes the beam-splitter scattering matrix, as
we shall see next.

\emph{Scattering approach.} After leaving the Rashba region of length $L$
within lead 1 an electron is left in a linear superposition of spin-up and
spin-down states with a phase $\theta _{R}$, Eq.~(\ref{eq5}). Both
components of this superposition freely and independently move into the
beam-splitter region where we assume they are partially transmitted to lead
3 or 4 . We can now combine the scattering matrices for the transmission of
electrons through the Rashba region $U_{R}$ and for the spin-independent
scattering of electrons from the incoming lead $\beta $ to the outgoing lead
$\alpha $ at the beam splitter $\mathbf{s}_{\alpha \beta }$. The combined
scattering matrices $\mathbf{s}_{31}^{R}=\mathbf{s}_{31}U_{R}=rU_{R} $ and $%
\mathbf{s}_{41}^{R}=\mathbf{s}_{41}U_{R}=tU_{R}$ describe both the Rashba
evolution within lead 1 and the subsequent transmission into leads 3 and 4.
Here, $r$ and $t$ denote the reflection and transmission amplitudes at the
beamsplitter, respectively. Electrons from lead 2 are similarly transmitted
into lead 3 or 4, but with no Rashba rotation. Here we have $\mathbf{s}_{23}=%
\mathbf{s}_{14}$ and $\mathbf{s}_{24}=\mathbf{s}_{13}$. Hence we define the
total scattering matrix
\begin{equation}
\mathbf{s}=\left(
\begin{array}{cccc}
0 & 0 & \mathbf{s}_{13}^{R} & \mathbf{s}_{14}^{R} \\
0 & 0 & \mathbf{s}_{23} & \mathbf{s}_{24} \\
\mathbf{s}_{31}^{R} & \mathbf{s}_{32} & 0 & 0 \\
\mathbf{s}_{41}^{R} & \mathbf{s}_{42} & 0 & 0%
\end{array}%
\right) ,  \label{eq10}
\end{equation}%
relevant for calculating transport properties. 

\emph{Shot noise.} Shot noise \cite{blanter} is a non-equilibrium current
fluctuation arising from the discrete nature of the charge flow (at zero
temperature). At a time $t$, the current fluctuation about its average in
lead $\alpha $ is $\delta \hat{I}_{\alpha }(t)=$ $\hat{I}_{\alpha
}(t)-\langle \hat{I}_{\alpha }\rangle $. As usual, shot noise is defined as
the Fourier transform of the symmetrized current-current autocorrelation
function between leads $\gamma $ and $\mu $
\begin{equation}
S_{\gamma \mu }(\omega )=\frac{1}{2}\int \langle \delta \hat{I}_{\gamma
}(t)\delta \hat{I}_{\mu }(t^{\prime })+\delta \hat{I}_{\mu }(t^{\prime
})\delta \hat{I}_{\gamma }(t)\rangle e^{i\omega t}dt.  \label{eq11}
\end{equation}%
The current in lead $\gamma $ in the scattering approach \cite{buttiker} is
\begin{eqnarray}
\hat{I}_{\gamma }(t) &=&\frac{e}{h}\sum_{\alpha \beta }\!\!\int
\!\!d\varepsilon d\varepsilon ^{\prime }e^{i(\varepsilon -\varepsilon
^{\prime })t/\hbar }\mathbf{a}_{\alpha }^{\dagger }(\varepsilon )\mathbf{A}%
_{\alpha \beta }(\gamma ;\varepsilon ,\varepsilon ^{\prime })\mathbf{a}%
_{\beta }(\varepsilon ^{\prime }),  \notag \\
&&\mathbf{A}_{\alpha \beta }(\gamma ;\varepsilon ,\varepsilon ^{\prime
})=\delta _{\gamma \alpha }\delta _{\gamma \beta }\mathbf{1}-\mathbf{s}%
_{\gamma \alpha }^{\dagger }(\varepsilon )\mathbf{s}_{\gamma \beta
}(\varepsilon ^{\prime }),
\end{eqnarray}%
where $\mathbf{a}_{\alpha}^\dagger = (a_{\alpha\uparrow}^\dagger,
a_{\alpha\downarrow}^\dagger)$, and $a_{\alpha \sigma }^{\dagger
}(\varepsilon)$ $[a_{\alpha \sigma }(\varepsilon)]$ denotes the creation
(annihilation) fermionic operator for an electron with energy $\varepsilon$
in lead $\alpha $; $\sigma $ is the spin component along a proper
quantization direction ($i=x,y,z$). The \emph{spin-dependent} \textbf{s}
matrix is defined in Eq.~(\ref{eq10}). Below we determine explicit formulas
for spin-polarized and spin-entangled electrons.

\emph{Spin-polarized electrons}. For Fermi liquid leads we obtain the
well-known noise formula \cite{buttiker} -- but with a spin-dependent
\textbf{s} matrix -- after performing the ensemble average $\langle \cdots
\rangle $ in Eq.~(\ref{eq11}). For spin-polarized electrons and a small bias
$eV$ applied between the incoming (1,2) and outgoing (3,4) leads we find to
linear order in $eV$ and at zero temperature
\begin{equation}
S_{33}^{p} = 2e I \, T(1-T) f_{p}, \quad f_{p} = p \,\sin^{2}\frac{\theta
_{R}}{2},  \label{eq13}
\end{equation}
where $T\equiv \left| t\right| ^{2}$ is the beam-splitter
transmission, $p$ the degree of spin polarization in leads 1 and
2, and $I=(2e^2/h) V /(1+p)$ the mean current in lead 3.

\emph{Spin-entangled electrons.} For electron pairs, the average in Eq.~(\ref%
{eq11}) is a quantum mechanical expectation value between two-electron
states. We consider the following injected states \cite{delay} in leads 1
and 2
\begin{eqnarray}
\left.
\begin{array}{c}
|S\rangle \\
|Te_{i}\rangle%
\end{array}%
\right\} &=&\frac{1}{\sqrt{2}}\left[ a_{1\uparrow }^{\dagger }(\varepsilon
_{1})a_{2\downarrow }^{\dagger }(\varepsilon _{2})\mp a_{1\downarrow
}^{\dagger }(\varepsilon _{1})a_{2\uparrow }^{\dagger }(\varepsilon _{2})%
\right] |0\rangle \text{,}  \notag \\
|Tu_{i}\rangle &=&a_{1\sigma }^{\dagger }(\varepsilon _{1})a_{2\sigma
}^{\dagger }(\varepsilon _{2})|0\rangle ,\text{ }\sigma =\uparrow
,\downarrow \text{,}  \label{eq14}
\end{eqnarray}%
where $|0\rangle $ denotes the ground state (filled or not) Fermi
sea of the leads. The states $|S\rangle $ and $|Te_{i}\rangle $
are the \emph{entangled}
singlet and triplets, respectively, while $|Tu_{i}\rangle $ are \emph{%
unentangled} triplets. Here we have in mind an entangler \cite{entangler}
attached to leads 1 and 2, Fig.~\ref{fig:fig1}. We assume that these pairs
have \emph{discrete} energies above $\varepsilon _{F}$ \cite{BLS}.

At zero temperature and applied voltage and $\omega =0$, the Fermi sea is
completely inert and the noise in the system is solely due to the injected
pairs above the Fermi surface \cite{BLS}. For the singlet and triplets in
Eq.~(\ref{eq14}) and the \textbf{s} matrix in Eq.~(\ref{eq10}), we find \cite%
{formula}
\begin{equation}
S_{33}^{X}=\frac{2e^{2}}{h\nu }T(1-T)f_{X},  \label{noise}
\end{equation}%
where the factor $f_{X}$, $X=S$, $Te_{i}$, $Tu_{i}$, $i=x,y,z$, depends on
the Rashba phase $\theta _{R}$. The density of states $\nu $ in (\ref{noise}%
) arises because of the discrete levels. For the spin singlet
\begin{equation}
f_{S}=1+\cos \theta _{R}\delta _{\varepsilon _{1}\varepsilon _{2}},
\label{f-singlet}
\end{equation}%
where $\varepsilon _{1}$ are $\varepsilon _{2}$ denote the discrete energies
of the paired electrons. For the triplet states in $y$ and $z$ directions
\begin{eqnarray}
f_{Te_{y}} &=&1-\cos \theta _{R}\delta _{\varepsilon _{1}\varepsilon _{2}},
\label{f-triplet-ey} \\
f_{Tu_{z}} &=&1-\cos ^{2}(\theta _{R}/2)\delta _{\varepsilon _{1}\varepsilon
_{2}},  \label{f-triplet-uz} \\
f_{Tu_{y}} &=&f_{Te_{z}}=1-\delta _{\varepsilon _{1}\varepsilon _{2}}.
\label{f-triplet-ez}
\end{eqnarray}

\emph{Two channels and s-o interband coupling.} So far we have considered a
strictly 1D lead with a local Rashba interaction. Now we consider the case
in which lead 1 has \emph{two} transverse Rashba channels $|a\rangle $ and $%
|b\rangle $ \cite{moroz}. We assume a weak s-o interband coupling which
splits the bands near the crossing point $k_{c}$, Fig.~\ref{fig:fig1}
(right). To lowest order this splitting is $2\alpha d$, $d\equiv \langle
a|d/dy|b\rangle $.

After traversing the Rashba region, a spin up (down) electron
impinging at the band crossing is left\ in the state
\begin{eqnarray}
&&e^{in\pi /2}\left(
\begin{array}{c}
e^{-i\theta _{R}/2}\cos \left( \theta _{d}/2\right) \pm e^{i\theta _{R}/2}
\\
-i\left( e^{-i\theta _{R}/2}\cos \left( \theta _{d}/2\right) \mp e^{i\theta
_{R}/2}\right)
\end{array}%
\right) \frac{|a\rangle }{2}  \notag \\
&&+e^{in\pi /2}\left(
\begin{array}{c}
-ie^{-i\theta _{R}/2}\sin \left( \theta _{d}/2\right)  \\
e^{-i\theta _{R}/2}\sin \left( \theta _{d}/2\right)
\end{array}%
\right) \frac{|b\rangle }{2},  \label{eq19a}
\end{eqnarray}%
where $\theta _{d}=\theta _{R}d/k_{c}$ \cite{formula}. To obtain
(\ref{eq19a}) we have
expanded an incoming spin-up $n=0$ (down, $n=1$) state in channel \emph{a }%
in terms of the s-o interband-coupled states near the energy crossing at $%
k_{c}$, see Fig. 1. We describe these interband-coupled states
perturbatively in analogy to the standard nearly-free electron model.
Incoming electrons are now injected into linear combinations of unperturbed
states (of channels \emph{a} and \emph{b}) near $k_{c}$ which satisfy proper
boundary conditions for the velocity operator \cite{molenkamp}.

Equation (\ref{eq19a}) clearly shows that impinging electrons with energies
near the band crossing undergo further spin rotation $\theta _{d}$. This
extra modulation arises because of channel mixing due to Rashba interband
coupling. For $\theta _{d}=0$, Eq. (\ref{eq19a}) yields the state (\ref{eq5}%
) with a single rotation $\theta _{R}$. An estimate of $\theta _{d}$ is
readily obtained for infinite transverse confinement: assuming an energy (at
the crossing)$\;\varepsilon _{F}=\varepsilon (k_{c})=24\epsilon _{R}=24\hbar
^{2}k_{R}^{2}/2m^{\ast }\Rightarrow \alpha d/\varepsilon (k_{c})\simeq 1/6$
\ and $d/k_{c}\simeq 0.5$; hence $\theta _{d}=\theta _{R}/2$ ($\theta
_{R}=\pi $ for $L=69$ nm and $\alpha =3.45\times 10^{-11}$ eVm \cite%
{nitta,grundler}; for this $\alpha $ the lateral width of the channel is $%
w=60$ nm). Therefore even a ``weak'' interband coupling yields a sizable
additional rotation $\theta _{d}$. This spin rotation for electrons injected
at the band crossing produces an additional modulation of the transport
properties. In particular, we find \cite{formula} for the spin-resolved
charge current in lead 1
\begin{equation}
I_{\uparrow ,\downarrow }\varpropto 1\pm \cos (\theta _{d}/2)\cos \theta
_{R},  \label{eq19b}
\end{equation}%
which clearly shows the additional modulation $\theta _{d}$ \cite{hausler}.

\emph{Generalized Fano factors.} To determine shot noise in the presence of
s-o interband coupling we proceed as before with the following extensions.
For electron pairs, for instance, we consider the states in Eq. (\ref{eq14}%
); here, however, the electron pair component in lead 1 evolves
according to (\ref{eq19a}). After a somewhat lengthy calculation
\cite{formula}, we find that the Fano factors for the noise are
now functions of both the Rashba angle $\theta _{R}$ and the
interband mixing angle $\theta _{d}$,
\begin{eqnarray}
f_{p} &=& \frac{p}{2}\! \left( 1-\cos \frac{\theta _{d}}{2}\cos \theta _{R}
+ \frac{1}{2}\sin^2\frac{\theta_d}{2}\right) , \hspace{18mm}  \label{eq13a}
\\
f_{S} &=& 1+\left( \cos \frac{\theta _{d}}{2}\cos \theta _{R}\right) \delta
_{\varepsilon _{1}\varepsilon _{2}},  \label{eq13aa} \\
f_{Te_{z}} &=& 1-\frac{1}{2}\! \left( \cos ^{2}\frac{\theta _{d}}{2}%
+1\right) \delta _{\varepsilon _{1}\varepsilon _{2}},  \label{eq13b} \\
f_{Tu_{z}} &=& 1 - \frac{1}{2}\! \left(\! 1+\cos \frac{\theta _{d}}{2}\cos
\theta _{R} - \frac{1}{2} \sin^2\frac{\theta_d}{2} \right) \!\delta
_{\varepsilon _{1}\varepsilon _{2}}.  \label{eq13c}
\end{eqnarray}%
The above equations reduce to the 1D case [Eqs. (\ref{eq13}) and (\ref%
{f-singlet})--(\ref{f-triplet-ez})] for $\theta _{d}=0$.

\begin{figure}[th]
\begin{center}
\epsfig{file=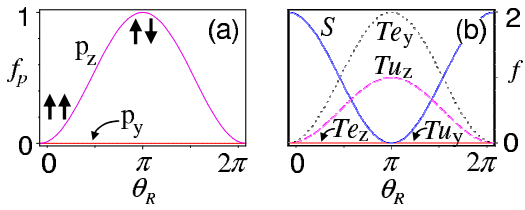, width=0.56\textwidth}
\end{center}
\par
\caption{Fano factor $f$ as a function of the Rashba phase $%
\protect\theta _{R}=2k_{R}L$ for (a) spin-polarized ($p=1$)
electrons and (b) triplet/singlet pairs. Spin-polarized electrons
exhibit nonzero noise only for the $z$ polarization; $y$-polarized
beams (along the Rashba rotation axis) yield $f=0$. At
$\protect\theta _{R}=\protect\pi $, full shot noise\ ($f=1$) is
recovered for the $z$-polarized beams. The entangled singlet state
$|S\rangle $ yields nonzero identical $f$'s for all quantization
axes. The triplets $|Te_{y}\rangle $ and $|Tu_{z}\rangle $ are
noisy while the triplets $|Tu_{y}\rangle $ and $|Te_{z}\rangle $
are noiseless, $f=0$. \ (b) Triplets and singlet show intermediate
degrees of bunching/antibunching.} \label{fig:fig2}
\end{figure}

\emph{Discussion. } Figure~\ref{fig:fig2} displays the ``normalized'' Fano
factor $f\equiv F/T(1-T)$, $F=S/2eI$, as a function of the Rashba phase $%
\theta _{R}=2k_{R}L$ for the (a) spin-polarized case [Eq. (\ref{eq13}) with $%
p=1$, here $I=e^{2}V/h$], (b) injected singlet and triplet pairs [Eqs. (\ref%
{f-singlet})--(\ref{f-triplet-ez}), here $I=e/h\nu $]. In Fig. 2
we plot $f$ for two quantization directions: $y$ and $z$ ($x$ is
equivalent to $z$), where ``$-y$'' defines the Rashba rotation
axis. In Fig.~\ref{fig:fig2}(a) only $z$-polarized electrons in
leads 1 and 2 generate noise as $\theta _{R}$ is varied;
$y$-polarized electrons are not affected by the Rashba rotation
about $-y$.

Because of the distinct symmetry of the orbital part of the pair wave
function, shot noise for singlet and triplet states is\emph{\ not} the same.
As detailed in Ref.~\cite{BLS} singlet pairs have a symmetric orbital wave
function thus showing ``bunching'' behavior; triplets, on the other hand,
show ``antibunching'' since their orbital wave function is antisymmetric.
The Rashba phase modifies the symmetry of the spin part of the pair
wavefunction; hence intermediate degrees of bunching or antibunching can be
induced.

Figure \ref{fig:fig2}(b) shows that shot noise for \emph{entangled} singlet
and triplet states display oscillatory bunching/antibunching behavior as a
function of the Rashba phase (singlet and triplets differ by $\pi $%
). Via these oscillations it is possible to distinguish the entangled
triplet states $|Te_{y}\rangle $ or $|Te_{z}\rangle $ from the respective
unentangled ones $|Tu_{y}\rangle $ or $|Tu_{z}\rangle $. For $%
\theta _{R}=0$ the difference in noise vanishes, i.e.,
$f_{Te_{i}}=f_{Tu_{i}} $\cite{BLS}.
\begin{figure}[th]
\begin{center}
\epsfig{file=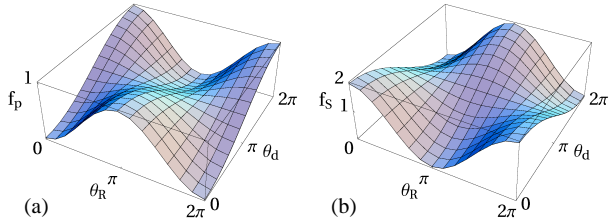,width=0.45\textwidth}
\end{center}
\par
\vspace{-4mm} \caption{Fano factor (a) $f_{p}$ for electrons
spin-polarized
along $\protect\sigma _{z}$ and (b) $f_{S}$ for spin singlets $%
|S\rangle $, as a function of $\protect\theta _{R}$ and $\protect\theta _{d}$%
. The additional phase $\protect\theta _{d}$ due to s-o interband coupling
allows for extra tuning of current and noise.}
\label{fig:fig3}
\end{figure}

Moreover, the oscillations in Fig.~\ref{fig:fig2} suggest a direct
way to obtain the s-o coupling constant $\alpha$ via measuring
shot noise; for instance, from Eq.~(\ref{eq13}) ($p=1$) we find
\begin{equation}
\alpha =\frac{\hbar ^{2}}{m^{\ast }L}\arcsin \sqrt{f_{p}}.  \label{eq20}
\end{equation}

Figure \ref{fig:fig3} illustrates the effect of the additional spin-rotation
$\theta _{d}$ (interband coupling) on the normalized Fano factor \emph{f};
only the spin-polarized and the singlet cases are shown. This extra rotation
can lead to a complete reversal of bunching/antibunching behavior for
electrons near the band crossing [see Fig. 1 (right)]. Hence additional spin
control is attained by varying the s-o interband coupling through the
lateral width of the confining potential (e.g., via side gates).

\emph{Conclusion.} Rashba s-o (interband) coupling strongly
modulates current and shot noise for spin-polarized and entangled
electrons in a beam-splitter geometry. This provides a means of
probing spin properties in charge transport and offers a direct
way to measure \textit{s-o} coupling constants.

This work was supported by NCCR\ Nanoscience, the Swiss NSF,
DARPA, and ARO. We acknowledge useful discussions with C. Schroll,
H. Gassmann, and D. Saraga.


\begin{thebibliography}{99}
\bibitem{dynamical} We can view the Rashba rotation (\ref{eq2}) as a time
evolution $U_{R}=\exp (-i\tau H_{R})$, within the electron traversal time $%
\tau =L/v_{F}$ ($v_{F}=\hbar k_{F}/m^{\ast }$) across the Rashba region.
\bibitem{mismatch} A very small mismatch arises because $\varepsilon =\hbar
^{2}(k\pm k_{R})^{2}/2m^{\ast }-\epsilon _{R}$ in the Rashba
region. To lowest order in the s-o coupling we find
$1-|r_{R}|^{2}\lesssim T_{R}\leqslant 1$, where
$r_{R}=[1-(1+\epsilon _{R}/\varepsilon
_{F})^{1/2}]/[1+(1+\varepsilon _{R}/\varepsilon _{F})^{1/2}]$ is
the reflection amplitude (single Rashba/normal interface)
\cite{molenkamp}, $\varepsilon _{F}$ is the Fermi energy, and
$\epsilon _{R}=\hbar ^{2}k_{R}^{2}/2m^{\ast }$ (Rashba energy).
For typical parameters $\epsilon _{R}\ll \varepsilon _{F}$,
e.g., $\varepsilon _{R}/\varepsilon _{F}=0.1$ \cite{nitta,grundler} we find $%
|r_{R}|^{2}<10^{-3}$.
\bibitem{delay} We neglect the ``delay time'' $\tau _{d}$ between the
injected partners in an electron pair. This is a good approximation for $%
\tau _{d}$ much smaller than the electron ``transit time'' $\tau _{t}\sim
L_{0}/v_{F}$ ($L_{0}$: ``device size''). For a superconductor-dot entangler %
\cite{entangler}, $\tau _{d}\sim \hbar /\Delta $ ($\Delta $: superconductor
gap); we then find $\tau _{d}\approx 0.6$ ps and $\tau _{t}\approx 100-10$
ps for $\Delta \approx 1$ meV, $v_{F}\approx 10^{4}-10^{5}$ m/s, and $%
L_{0}\approx 1$ $\mu $m.
\end{thebibliography}
\end{document}